\documentclass[aps,prb,twocolumn,floatfix,amsmath,amssymb,superscriptaddress]{revtex4-2}
\usepackage{graphicx}
\usepackage[T1]{fontenc}

\newcommand{\BFCA}{Ba(Fe$_{1-x}$Co$_x$)$_2$As$_2$}

\newcommand{\FS}{FeSe}
\newcommand{\FT}{Fe$_{1.07}$Te}
\newcommand{\BKFA}{Ba$_{0.83}$K$_{0.17}$Fe$_2$As$_2$}

\begin{document}

\title{$d$-Spacing Distributions as a Probe of Nematoelastic Response in Iron-Based Superconductors}

\author{Wenting Zhang}
\author{Ruixian Liu}
\email{liurx@mail.bnu.edu.cn}
\affiliation{School of Physics and Astronomy, Beijing Normal University, and Key Laboratory of Multiscale Spin Physics (Beijing Normal University), Ministry of Education, Beijing 100875, China}
\author{Tingjun Zhang}
\affiliation{Department of Physics and Astronomy, Rice University, Houston, Texas 77005, USA}
\affiliation{Rice Laboratory for Emergent Magnetic Materials and Smalley-Curl Institute, Rice University, Houston, Texas 77005, USA}
\affiliation{Applied Physics Graduate Program, Smalley-Curl Institute, Rice University, Houston, Texas 77005, USA}

\author{Weiliang Yao}
\affiliation{Department of Physics and Astronomy, Rice University, Houston, Texas 77005, USA}

\author{X\"ue Fu}
\author{Hanqing Xie}
\author{Ziye Mo}
\author{Ting Guo}
\affiliation{School of Physics and Astronomy, Beijing Normal University, and Key Laboratory of Multiscale Spin Physics (Beijing Normal University), Ministry of Education, Beijing 100875, China}
\author{Kuo-Feng Tseng}
\author{Thomas Keller}
\affiliation{Max Planck Society Outstation at the Forschungsneutronenquelle Heinz Maier-Leibnitz (MLZ), D-85747 Garching, Germany}
\author{Jitae T. Park}
\affiliation{Heinz Maier-Leibnitz Zentrum (FRM-II), Technische Universit\"at M\"unchen, 85748 Garching, Germany}

\author{Fankang Li}
\affiliation{Neutron Sciences Directorate, Oak Ridge National Laboratory, Oak Ridge, TN 37830, USA}

\author{Masaaki Matsuda}
\affiliation{Neutron Sciences Directorate, Oak Ridge National Laboratory, Oak Ridge, TN 37830, USA}

\author{Avishek Maity}
\affiliation{Neutron Sciences Directorate, Oak Ridge National Laboratory, Oak Ridge, TN 37830, USA}
\author{Long Tian}
\affiliation{School of Physics and Astronomy, Beijing Normal University, and Key Laboratory of Multiscale Spin Physics (Beijing Normal University), Ministry of Education, Beijing 100875, China}

\author{Pengcheng Dai}
\email{pdai@rice.edu}
\affiliation{Department of Physics and Astronomy, Rice University, Houston, Texas 77005, USA}
\affiliation{Rice Laboratory for Emergent Magnetic Materials and Smalley-Curl Institute, Rice University, Houston, Texas 77005, USA}

\author{Xingye Lu}
\email{luxy@bnu.edu.cn}
\affiliation{School of Physics and Astronomy, Beijing Normal University, and Key Laboratory of Multiscale Spin Physics (Beijing Normal University), Ministry of Education, Beijing 100875, China}

\date{\today}

\begin{abstract}

Electronic nematicity in iron-based superconductors (FeSCs) couples bilinearly to orthorhombic strain, allowing nematic correlations to appear in the lattice response. Here we use neutron Larmor diffraction to measure the temperature-dependent distribution of relative $d$ spacings in electron-doped {\BFCA}, hole-doped {\BKFA}, {\FS}, and {\FT}. In {\BFCA} crystals without intentionally applied uniaxial stress, the in-plane distribution width, $\varepsilon_{\rm FWHM}$, increases on cooling in the tetragonal phase and can be described phenomenologically by a Curie--Weiss-like form. The fitted scale $T^*$ decreases with Co doping and evolves similarly to the nematic phase diagram inferred from elastoresistance, although the two experiments probe different response functions. Related broadening in {\BKFA} and {\FS} supports extending this interpretation beyond electron-doped BaFe$_2$As$_2$. By contrast, {\FT} shows no extended Curie--Weiss-like regime without applied stress, whereas uniaxial pressure produces a strongly anisotropic broadening that can contain contributions from both the field-biased lattice response and inhomogeneous loading. A mean-field model with bilinear nematoelastic coupling and spatially varying symmetry-breaking stress explains the Curie--Weiss-like broadening in terms of the renormalized orthorhombic compliance. Neutron Larmor diffraction therefore provides a bulk-sensitive probe of nematic-related lattice broadening that complements electronic and elastic measurements.

\end{abstract}

\maketitle

\section{Introduction}
\vspace{-0.8em}

Electronic nematicity is one of the most prominent intertwined electronic orders in iron-based superconductors (FeSCs), characterized by the spontaneous breaking of the tetragonal $C_4$ rotational symmetry to a twofold $C_2$ symmetry without changing the translational symmetry of the lattice \cite{fernandes2014what,bohmer2022nematicity,si2016high,lu2026iron}. Although the accompanying crystallographic orthorhombicity is typically very small ($\sim 10^{-3}$), the electronic anisotropies associated with the nematic state can be remarkably large, indicating that the broken rotational symmetry is primarily electronic in origin \cite{chu2012divergent}. Experimentally, electronic nematicity manifests itself in a wide range of physical properties \cite{bohmer2022nematicity,lu2026iron}. Early transport measurements on detwinned crystals revealed pronounced in-plane resistivity anisotropy, while elastoresistivity measurements established a Curie--Weiss divergence of the nematic susceptibility in the tetragonal phase \cite{chu2010,chu2012,kuo2016ubiquitous}. Momentum-resolved spectroscopies have shown orbital-dependent band reconstruction and the lifting or rearrangement of the near degeneracy between Fe $d_{xz}$ and $d_{yz}$ orbitals \cite{yi2011symmetry,pfau2019momentum,yi2019nematic}. Nematicity also leaves strong fingerprints in spin dynamics, including twofold anisotropic spin excitations detected by inelastic neutron scattering and resonant inelastic x-ray scattering \cite{lu2014nematic,lu2018spin,chen2019anisotropic,tam2020orbital,lu2022spinexcitation,liu2024nematic,liu2025spin}. Complementary local and bulk probes, including STM, NMR, and Raman scattering, have further revealed $C_2$-symmetric electronic textures, anisotropic spin-lattice relaxation, and symmetry-resolved nematic fluctuations \cite{gallais2016nematic,thorsmolle2016critical,kissikov2016nmr,li2017stripes}.

Several of these signatures are coupled to the lattice through the nematoelastic interaction. Because the electronic nematic order parameter and the corresponding orthorhombic shear strain transform in the same symmetry channel, a bilinear coupling is allowed in the Landau free energy \cite{fernandes2014what,bohmer2022nematicity}. The coupling produces a tetragonal-to-orthorhombic distortion, makes symmetry-breaking strain a conjugate field to nematic order, and softens the shear modulus $C_{66}$ \cite{fernandes2010effects,yoshizawa2012structural,bohmer2014nematic,bohmer2016electronic,fujii2018anisotropic}. Related signatures occur in transverse acoustic phonons and in the strain dependence of electronic and magnetic transitions \cite{PhysRevB.91.134426,PhysRevX.8.021056,worasaran2021nematic}. The lattice response is not confined to the long-range orthorhombic phase below $T_s$: neutron Larmor diffraction detects anomalous $d$-spacing broadening, pair-distribution-function measurements resolve short-range orthorhombicity, and dark-field x-ray microscopy images mesoscopic shear-strain textures in nominally tetragonal or spatially heterogeneous regimes \cite{lu2016impact,wang2018local,frandsen2017local,frandsen2019quantitative,yay2026discovery}.

Previous neutron Larmor diffraction studies have established that the $d$-spacing distribution is closely tied to the coupled nematic--lattice response in FeSCs. In BaFe$_2$As$_2$-based compounds, uniaxial pressure acts as a conjugate field to electronic nematic order, inducing finite orthorhombicity and producing Curie--Weiss-like broadening of the in-plane $d$-spacing distribution \cite{lu2016impact,man2015electronic}. The same work also showed that BaFe$_{1.97}$Ni$_{0.03}$As$_2$ exhibits Curie--Weiss-like broadening without applied pressure, indicating that local lattice broadening can persist without an intentionally applied symmetry-breaking field \cite{lu2016impact}. These results provided early evidence that neutron Larmor diffraction is sensitive not only to the average orthorhombic distortion, but also to spatially distributed nematic lattice responses.
This picture was further developed by neutron Larmor diffraction measurements on NaFe$_{1-x}$Ni$_x$As, where $\varepsilon_{\rm FWHM}$ of the relative $d$-spacing distribution exhibits a Curie--Weiss-like temperature dependence in the nominally tetragonal phase without intentionally applied uniaxial stress \cite{wang2018local}. This behavior was attributed to weak local symmetry-breaking fields, originating from quenched disorder or residual strain, that are amplified by the soft nematoelastic response and produce quasi-static local orthorhombic distortions with zero macroscopic average. These observations suggest that neutron Larmor diffraction can detect nematic-related lattice broadening even when the average structure remains tetragonal.

Despite these advances, several central questions remain unresolved. It is still unclear whether Curie--Weiss-like broadening of the $d$-spacing distribution in the nominally unpressured state is a common feature of FeSCs or is limited to specific materials and disorder environments. It is also essential to clarify how $\varepsilon_{\rm FWHM}$ should be interpreted under external uniaxial stress, where the applied load can simultaneously generate average orthorhombicity and an inhomogeneous local strain distribution. This issue is particularly important for materials with strongly first-order structural or magnetostructural transitions. In SrFe$_{1.97}$Ni$_{0.03}$As$_2$, the unpressured $d$-spacing width shows no extended Curie--Weiss-like enhancement above the first-order transition, whereas a pronounced Curie--Weiss-like broadening emerges under uniaxial pressure \cite{lu2016impact}. A systematic comparison across different FeSC families and loading conditions is therefore required to establish neutron Larmor diffraction as a bulk probe of nematoelastic lattice response and to separate intrinsic local orthorhombic broadening from pressure-induced strain inhomogeneity.

\begin{figure}[t]
\includegraphics[width=8.5cm]{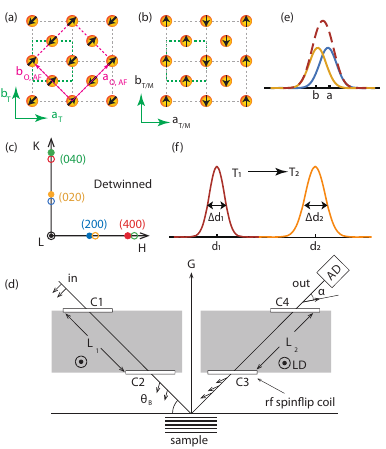}
\caption{Crystal and magnetic structures, reciprocal-space configuration, and neutron Larmor diffraction principle. 
(a) Schematic in-plane structure and stripe-type antiferromagnetic spin arrangement of BaFe$_2$As$_2$-based compounds. The tetragonal axes $a_T$ and $b_T$ and the orthorhombic/magnetic axes $a_{O,\mathrm{AF}}$ and $b_{O,\mathrm{AF}}$ are indicated. 
(b) Schematic in-plane structure and bicollinear antiferromagnetic spin arrangement of {\FT}. The tetragonal/monoclinic axes $a_{T/M}$ and $b_{T/M}$ are indicated. 
(c) Reciprocal-space configuration for a detwinned sample, illustrating the separation of symmetry-related in-plane Bragg reflections, such as $(4,0,0)$ and $(0,4,0)$ for Ba122 compounds and $(2,0,0)$ and $(0,2,0)$ for iron chalcogenides. 
(d) Schematic neutron Larmor diffraction setup. Polarized neutrons acquire Larmor phases before and after Bragg scattering, allowing the lattice spacing selected by a Bragg reflection to be encoded in the neutron polarization. 
(e) Schematic diffraction response associated with orthorhombic splitting of in-plane $d$-spacings.
(f) Temperature evolution of the $d$-spacing distribution (from temperature $T_1$ to temperature $T_2$), illustrating changes in both the average lattice spacing and the distribution width $\Delta d$.}
\label{fig1}
\end{figure}

In this work, we use neutron Larmor diffraction to establish the relative $d$-spacing width $\varepsilon_{\rm FWHM}$ as a bulk-sensitive lattice probe of the nematoelastic response in FeSCs. We measure the temperature-dependent $d$-spacing distributions of electron-doped {\BFCA}, hole-doped {\BKFA}, {\FS}, and {\FT}, thereby covering systems with continuous or weakly first-order nematic transitions, a nematic chalcogenide without ambient-pressure long-range magnetic order, and a strongly first-order magnetostructural compound. In {\BFCA}, the in-plane $\varepsilon_{\rm FWHM}$ follows a Curie--Weiss-like temperature dependence above $T_s$, and the fitted scale $T^*$ decreases systematically with Co doping in close correspondence with the nematic phase diagram inferred from elastoresistance \cite{chu2012}. The observation of similar Curie--Weiss-like broadening in {\BKFA} and {\FS} shows that this lattice response is not restricted to electron-doped BaFe$_2$As$_2$, while the contrasting behavior of {\FT} demonstrates how a strongly first-order transition suppresses the strain-free Curie--Weiss-like broadening and how uniaxial pressure restores a pronounced, anisotropic response. We further develop a mean-field description in which local symmetry-breaking stresses are amplified by the soft orthorhombic compliance generated by nematoelastic coupling. This framework unifies the strain-free broadening observed in {\BFCA}, {\BKFA}, and {\FS} with the pressure-enhanced broadening in {\FT}, and provides a direct structural route for probing nematic-related lattice responses through the full $d$-spacing distribution.

\section{Experimental Method}

\begin{figure}[htbp!]
\includegraphics[width=8cm]{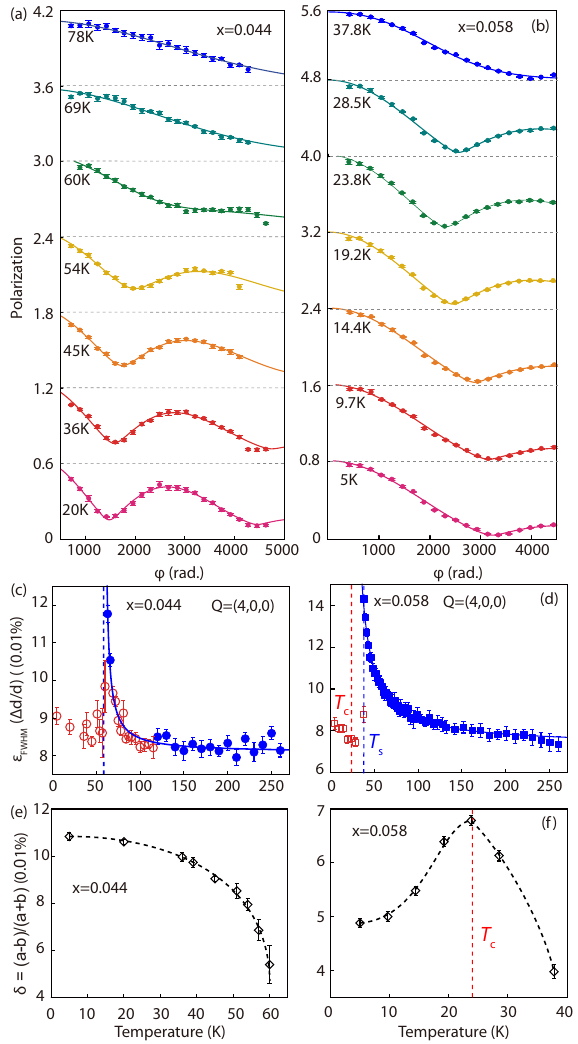}
\caption{
Neutron Larmor diffraction results for {\BFCA} without intentionally applied uniaxial stress, measured at the $(4,0,0)$ Bragg reflection.
(a),(b) Neutron polarization as a function of the total Larmor phase $\phi$ for samples with $x=0.044$ ($T_s\approx60$ K, $T_{c}\approx10$ K) and $x=0.058$ ($T_s\approx38$ K, $T_{c}\approx24$ K), respectively, at selected temperatures. Solid lines are fits to the polarization curves. The polarization curves are vertically offset for clarity; gray dashed lines indicate the corresponding baselines.
(c),(d) Temperature dependence of the $d$-spacing width 
$\varepsilon_{\rm FWHM}$ for $x=0.044$ and $x=0.058$, respectively. 
Solid lines are Curie--Weiss-like fits to the data in the tetragonal phase. 
Open symbols denote values extracted directly from the Larmor-diffraction polarization curves shown in (a) and (b), whereas filled symbols were obtained from thermal-expansion measurements as described in the text. 
Vertical blue and red dashed lines mark $T_s$ and $T_c$, respectively. 
(e),(f) Temperature dependence of the orthorhombic lattice distortion 
$\delta=(a-b)/(a+b)$ for $x=0.044$ and $x=0.058$, respectively, obtained from the resolved splitting of the in-plane lattice-spacing components. 
Dashed lines are guides to the eye.
}
\label{fig2}
\end{figure}

Single crystals of {\BFCA} with Co concentrations $x=0.034$, $0.044$, $0.058$, $0.0658$, and $0.0673$ \cite{tian2019spin}, together with {\BKFA} \cite{li2025neutron}, were grown by the self-flux method [Fig.~\ref{fig1}(a)]. Single crystals of {\FS} \cite{liu2024low} and {\FT} \cite{song2018spinisotropic} were prepared by chemical vapor transport and a modified Bridgman method, respectively. Measurements on {\BFCA} with $x=0.058$, $0.0658$, and $0.0673$, as well as on {\FT}, were performed on the three-axis spin-echo spectrometer TRISP at the Heinz Maier-Leibnitz Zentrum (MLZ), Garching, Germany \cite{keller2002,ono2002neutron}. Measurements on {\BFCA} with $x=0.034$ and $0.044$, {\BKFA}, and {\FS} were carried out on the polarized triple-axis spectrometer PTAX at the High Flux Isotope Reactor (HFIR), Oak Ridge National Laboratory (ORNL), USA \cite{li2017high,li2018new}. Unless stated otherwise, no intentional uniaxial load was applied; this condition does not exclude residual internal or mounting stress. {\FT} was also measured under a nominal uniaxial pressure $P>15$~MPa applied along the in-plane $b$ direction. The $(4,0,0)$ Bragg reflection was used for {\BFCA} and {\BKFA}, whereas the $(2,0,0)$ reflection was used for {\FS} and {\FT} [Fig.~\ref{fig1}(c)].

\begin{figure*}[htbp!]
\includegraphics[width=17 cm]{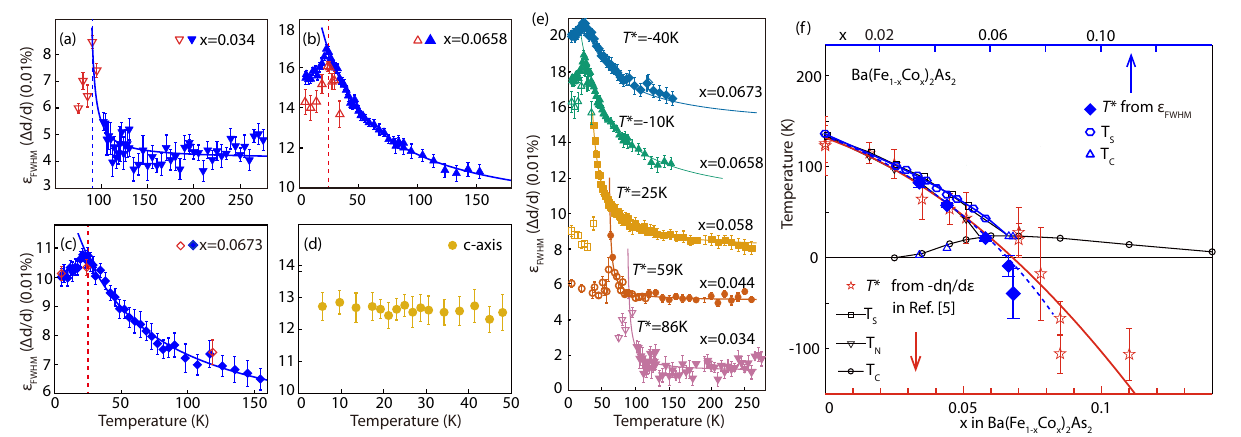}
\caption{
Doping evolution of the relative $d$-spacing width in {\BFCA} without intentionally applied uniaxial stress.
(a)--(c) Temperature dependence of $\varepsilon_{\rm FWHM}$ at the in-plane $(4,0,0)$ reflection for $x=0.034$ ($T_s\approx90$ K, $T_{c}\approx5$ K), $0.0658$ ($T_{c}\approx25.5$ K), and $0.0673$ ($T_{c}\approx25$ K), respectively. Solid lines are phenomenological Curie--Weiss-like fits in the tetragonal phase.
(d) Width measured along the $c$ axis, showing no comparable temperature-dependent anomaly.
(e) Comparison of the in-plane widths for $x=0.034$, $0.044$, $0.058$, $0.0658$, and $0.0673$. The labeled $T^*$ values are obtained from the same empirical fits. Open symbols in (a)-(e) denote values extracted from Larmor-diffraction polarization curves, whereas filled symbols were obtained from thermal-expansion measurements. 
(f) Larmor-diffraction $T^*$ values (blue diamonds) compared with the {\BFCA} phase diagram inferred from constant-strain elastoresistance ($-d\eta/d\varepsilon$) \cite{chu2012}. The two sets of scales arise from different observables and are compared here primarily through their doping evolution.
}
\label{fig3}
\end{figure*}

For a Bragg reflection indexed by $(H,K,L)$, the scattering vector is $\mathbf{Q}=H\mathbf{a}^{*}+K\mathbf{b}^{*}+L\mathbf{c}^{*}$, where $\mathbf{a}^{*}=2\pi\hat{\mathbf{a}}/a$, $\mathbf{b}^{*}=2\pi\hat{\mathbf{b}}/b$, and $\mathbf{c}^{*}=2\pi\hat{\mathbf{c}}/c$. The corresponding plane spacing is $d=2\pi/|\mathbf{Q}|$. In neutron Larmor diffraction, polarized neutrons acquire a Larmor precession phase before and after Bragg scattering [Fig.~\ref{fig1}(d)]. The accumulated phase is proportional to the $d$ spacing selected by the Bragg reflection, $\phi_{\rm tot}\propto d$, and a relative distribution of spacings produces $\Delta\phi_{\rm tot}=\phi_{\rm tot}\Delta d/d$. The phase dependence of the neutron polarization $P(\phi_{\rm tot})$ therefore encodes both the average spacing and its distribution [Figs.~\ref{fig1}(e) and \ref{fig1}(f)].

For a single Gaussian distribution, the polarization decay is
$P(\phi_{\rm tot})=P_0\exp[-\phi_{\rm tot}^2\varepsilon_{\rm FWHM}^2/(16\ln2)]$,
where $\varepsilon_{\rm FWHM}\equiv(\Delta d/d)_{\rm FWHM}$ is the full width at half maximum of the relative $d$-spacing distribution \cite{lu2016impact}. 
Representative full $P(\phi_{\rm tot})$ scans were measured at selected temperatures to verify the Gaussian line shape and to determine the normalization factor $P_0$. Representative full $P(\phi_{\rm tot})$ scans indicate that $P_0$ is nearly temperature independent in the tetragonal phase within experimental uncertainty. Therefore, most temperature-dependent measurements of $\varepsilon_{\rm FWHM}$ were obtained by measuring the polarization at a fixed value of $\phi_{\rm tot}$ (thermal expansion measurements) and solving the above expression for the width,
$\varepsilon_{\rm FWHM}=(4\sqrt{\ln2}/|\phi_{\rm tot}|)\sqrt{\ln[P_0/P(\phi_{\rm tot})]}$. 
This procedure provides an efficient way to track the temperature evolution of the $d$-spacing spread, while full $P(\phi_{\rm tot})$ scans were used to check the validity of the single-Gaussian description. When the orthorhombic splitting is resolved, the polarization curves are fitted with a two-component Gaussian model and the distortion is obtained from the component separation as $\delta=(a-b)/(a+b)=\Delta\varepsilon/2$ \cite{lu2016impact}.

In the tetragonal phase, where no orthorhombic splitting is resolved, the extracted $\varepsilon_{\rm FWHM}$ exhibits a Curie--Weiss-like temperature dependence and can be fitted by
$\varepsilon_{\rm FWHM}(T)=A/(T-T^*)+B$ \cite{lu2016impact,wang2018local}. 
Here $A$ and $B$ are fitting constants, and $T^*$ is the characteristic temperature of the Curie--Weiss-like lattice broadening. As discussed below, this behavior arises naturally when local symmetry-breaking fields are amplified by the soft nematoelastic response.

\section{Results}

We first focus on {\BFCA}, a canonical system in which electronic nematicity, orthorhombic lattice distortion, magnetism, and superconductivity have been extensively characterized. Previous neutron Larmor diffraction measurements under uniaxial pressure showed that the pressure-induced orthorhombicity and the in-plane $d$-spacing spread are strongly enhanced by the nematoelastic response \cite{lu2016impact}. Here we examine whether a similar nematic-related lattice broadening can be detected in crystals without intentionally applied uniaxial stress.

Figure~\ref{fig2} presents representative Larmor diffraction measurements on {\BFCA} with $x=0.044$ and $x=0.058$, measured at the in-plane $(4,0,0)$ Bragg reflection. For both compositions, the polarization curves $P(\phi)$ evolve strongly with temperature [Figs.~\ref{fig2}(a) and \ref{fig2}(b)], demonstrating that the underlying $d$-spacing distribution changes substantially on cooling. In the tetragonal phase, the polarization decay is well described by the single-Gaussian model, from which the relative $d$-spacing width $\varepsilon_{\rm FWHM}$ is extracted. Below the structural transition, the polarization curves are analyzed using the two-component model when the orthorhombic splitting becomes resolvable, allowing both the distribution width and the orthorhombic distortion to be determined.

The extracted $\varepsilon_{\rm FWHM}$ increases markedly on cooling toward $T_s$ from the tetragonal phase [Figs.~\ref{fig2}(c) and \ref{fig2}(d)]. Above $T_s$, the temperature dependence can be described by a Curie--Weiss-like form, indicating that the in-plane lattice broadening is governed by the enhanced nematoelastic response. Below $T_s$, the orthorhombic splitting of the in-plane lattice spacing becomes visible. For the underdoped $x=0.044$ sample, the orthorhombic distortion $\delta=(a-b)/(a+b)$ develops clearly below $T_s$ and increases on cooling [Fig.~\ref{fig2}(e)]. For the more highly doped $x=0.058$ sample, the distortion is much weaker and is suppressed below $T_c$ [Fig.~\ref{fig2}(f)], consistent with the competition between superconductivity and nematic/orthorhombic order \cite{wang2018local}.

To determine how this lattice response evolves with electron doping, we measured the in-plane $d$-spacing width over a broader range of Co concentrations, as summarized in Fig.~\ref{fig3}. The in-plane $\varepsilon_{\rm FWHM}$ shows pronounced temperature-dependent enhancement from the underdoped composition $x=0.034$ to the nearly overdoped composition $x=0.0673$ [Figs.~\ref{fig3}(a)--\ref{fig3}(c)]. Although the magnitude of the enhancement is gradually reduced with increasing Co concentration, the Curie--Weiss-like behavior remains clearly visible over a wide doping range. This systematic evolution is consistent with the progressive weakening of nematicity by electron doping.

The directional dependence provides an important control. As shown in Fig.~\ref{fig3}(d), the width measured along the $c$ axis exhibits only weak temperature dependence and does not show a comparable anomaly. This contrast demonstrates that the observed broadening is associated primarily with the in-plane symmetry-breaking lattice channel, rather than with a generic thermal, instrumental, or sample-wide broadening effect. Figure~\ref{fig3}(e) compares the in-plane $\varepsilon_{\rm FWHM}$ for all measured Co concentrations. Curie--Weiss-like fits to the tetragonal-region data yield a characteristic temperature $T^*$ that decreases systematically with increasing Co doping, evolving from positive values in the underdoped regime to negative values near the overdoped side. As shown in Fig.~\ref{fig3}(f), this doping evolution closely follows the nematic phase diagram obtained from elastoresistance measurements \cite{chu2012,kuo2016ubiquitous}. The close correspondence between these two independent measurements shows that the in-plane $d$-spacing distribution measured by neutron Larmor diffraction provides a bulk lattice measure of the nematoelastic response in nominally unstressed {\BFCA}.

\begin{figure}[b]
\includegraphics[width=8.5cm]{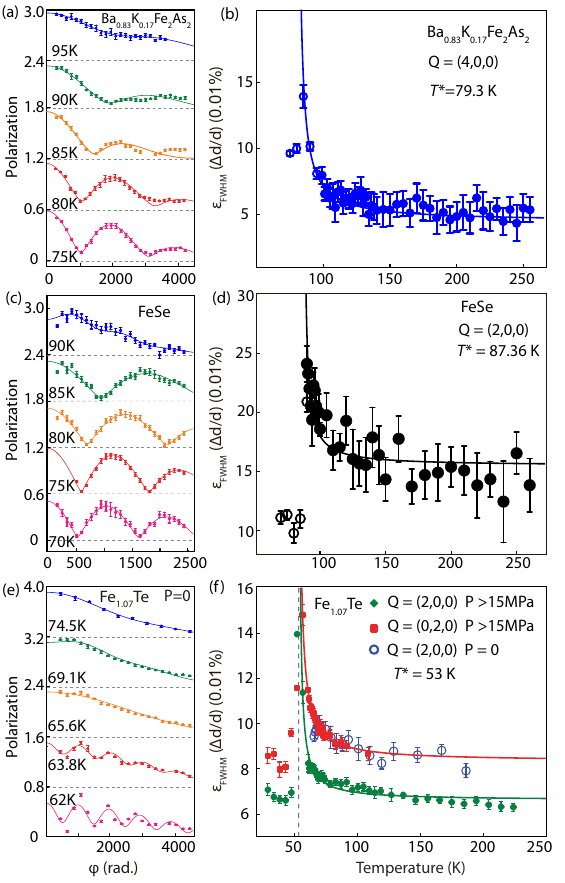}
\caption{
Neutron Larmor diffraction results for {\BKFA}, {\FS}, and {\FT}.
(a), (c), (e) Neutron polarization as a function of the total Larmor phase $\phi$ for {\BKFA}, {\FS}, and {\FT}, respectively. Solid lines are fits to the data.
(b),(d) Temperature dependence of $\varepsilon_{\rm FWHM}$ for {\BKFA} and {\FS}, respectively. Solid lines are Curie--Weiss-like fits, giving $T^*\approx79$~K for {\BKFA} and $T^*\approx87$~K for {\FS}.
(f) Widths for {\FT} at the $(2,0,0)$ and $(0,2,0)$ reflections under nominal uniaxial pressure $P>15$~MPa, compared with the $(2,0,0)$ result obtained without intentionally applied pressure. Open symbols in panels (b), (d), and (f) denote values extracted from the Larmor-diffraction measurements shown in panels (a), (c), and (e), respectively. Filled symbols denote values determined from thermal-expansion measurements.
}
\label{fig4}
\end{figure}

We next extend the measurements to {\BKFA} ($T_s=T_{N}\approx 103$ K), {\FS} ($T_s\approx 90$ K), and {\FT} ($T_s=T_N\approx 67$ K), as summarized in Fig.~\ref{fig4}. These materials provide three complementary comparisons with {\BFCA}. {\BKFA} tests whether the same behavior appears on the hole-doped side of the BaFe$_2$As$_2$ phase diagram. {\FS} provides a structurally simple iron chalcogenide in which nematic order develops without static long-range antiferromagnetic order at ambient pressure. {\FT} provides a contrasting case with a strongly first-order magnetostructural transition.

For hole-doped {\BKFA}, the polarization curves measured at the $(4,0,0)$ reflection show a clear temperature-dependent decay with increasing Larmor phase [Fig.~\ref{fig4}(a)], indicating a progressive broadening of the $d$-spacing distribution on cooling. The extracted $\varepsilon_{\rm FWHM}$ exhibits a Curie--Weiss-like enhancement in the tetragonal phase, yielding $T^*\approx79$~K [Fig.~\ref{fig4}(b)]. The observation of the same type of lattice broadening in hole-doped {\BKFA} shows that this effect is not restricted to electron-doped {\BFCA}, but is also present on the hole-doped side of the BaFe$_2$As$_2$-based phase diagram.

A similar response is observed in {\FS}. As shown in Fig.~\ref{fig4}(c), the polarization measured at the $(2,0,0)$ reflection evolves strongly with temperature and is well described by the same fitting procedure. The resulting $\varepsilon_{\rm FWHM}$ increases rapidly on cooling in the tetragonal phase and follows a Curie--Weiss-like temperature dependence, with $T^*\approx87$~K [Fig.~\ref{fig4}(d)]. This scale is close to the nematic transition temperature of FeSe and is consistent with transport-based measurements of the nematic response \cite{liu2025evolutionb}. Since FeSe lacks static long-range antiferromagnetic order at ambient pressure, the Curie--Weiss-like broadening observed here shows that the Larmor-diffraction width is sensitive to the nematic lattice channel itself and does not require static magnetic order.

{\FT} provides an important contrast to the continuous-transition systems. In the state without intentionally applied uniaxial pressure, the $d$-spacing width of {\FT} does not show an extended Curie--Weiss-like enhancement above the transition [Figs.~\ref{fig4}(e) and blue circles in \ref{fig4}(f)]. This is consistent with the strongly first-order nature of its magnetostructural transition \cite{li2009firstorder}, which cuts off the gradual nematoelastic softening observed in {\BFCA}, {\BKFA}, and {\FS}. Under applied uniaxial pressure, however, the in-plane $\varepsilon_{\rm FWHM}$ becomes strongly enhanced and develops a pronounced temperature dependence. Moreover, the broadening is larger for the reflection associated with the pressure-applied direction, indicating that the applied stress both biases the orthorhombic response and generates an anisotropic distribution of local strain. This behavior closely parallels the pressure-induced broadening previously observed in SrFe$_{1.97}$Ni$_{0.03}$As$_2$ \cite{lu2016impact}.

These measurements extend the conclusions drawn from {\BFCA} to a broader set of FeSCs. The Curie--Weiss-like enhancement of $\varepsilon_{\rm FWHM}$ in {\BKFA} and {\FS} supports the view that in-plane $d$-spacing broadening is a general lattice manifestation of enhanced nematoelastic response. The contrasting behavior of {\FT} further shows that the character of the structural transition and the presence of external uniaxial stress play decisive roles in shaping the measured distribution. Together, the results provide the experimental basis for a unified description of neutron Larmor diffraction width in terms of nematoelastic coupling, local symmetry-breaking fields, and strain inhomogeneity.

\section{Discussion}
\vspace{-0.8em}

The central experimental observation of this work is that the in-plane
relative $d$-spacing width, $\varepsilon_{\rm FWHM}$, develops a
Curie--Weiss-like temperature dependence in several FeSCs with
enhanced nematic response. This behavior can be understood within a
mean-field nematoelastic framework in which weak local
symmetry-breaking fields are amplified by the soft orthorhombic
elastic channel. We introduce an electronic nematic order parameter
$\phi$ and the orthorhombic strain $\varepsilon_{6}=(a-b)/(a+b)$.
These two quantities transform in the same orthorhombic shear channel
and are therefore coupled bilinearly. To describe the distribution
measured by Larmor diffraction, we further include a spatially varying
symmetry-breaking stress $\sigma_6(\mathbf r)$, which may originate
from quenched disorder, compositional variations, residual local
strain, or an inhomogeneously transmitted external load. The minimal
local free-energy density is
\begin{equation}
\begin{split}
f =&\,
\frac{a}{2}(T-T_{0})\phi^{2}
+\frac{u}{4}\phi^{4}
+\frac{C_{66,0}}{2}\varepsilon_{6}^{2}  \\
&-\lambda\phi\varepsilon_{6}
-\sigma_{6}(\mathbf{r})\varepsilon_{6},
\end{split}
\label{eq:landau_nematic}
\end{equation}
where $T_{0}$ is the bare electronic nematic Weiss temperature,
$C_{66,0}$ is the noncritical shear modulus, and $\lambda$ is the
nematoelastic coupling constant
\cite{fernandes2010effects,bohmer2014nematic,bohmer2016electronic}.
In the tetragonal phase, the bare electronic nematic susceptibility is
$\chi_{\phi}^{(0)}=[a(T-T_{0})]^{-1}$. The bilinear coupling
renormalizes the shear modulus according to
\begin{equation}
\begin{split}
C_{66}(T)
&= C_{66,0}-\lambda^{2}\chi_{\phi}^{(0)}(T)  \\
&= C_{66,0}
\frac{T-T_{s}^{\rm CW}}{T-T_{0}},
\end{split}
\label{eq:c66_cw}
\end{equation}
where $T_{s}^{\rm CW}=T_{0}+\lambda^{2}/(aC_{66,0})$ is the
Curie--Weiss scale of the coupled nematoelastic response. Thus,
nematic fluctuations soften the orthorhombic elastic channel and
convert local symmetry-breaking fields into an enhanced lattice
response.

For a nominally strain-free crystal, the spatial average of the local
stress can vanish, $\overline{\sigma_6(\mathbf r)}=0$, while its
variance remains finite, $\overline{\sigma_6^2(\mathbf r)}=\Delta_\sigma^2$.
Minimizing Eq.~(\ref{eq:landau_nematic}) gives
\begin{equation}
\varepsilon_{6}(\mathbf{r},T)
=
\frac{\sigma_{6}(\mathbf{r})}{C_{66}(T)} .
\label{eq:random_stress}
\end{equation}
The average structure can therefore remain tetragonal, even though
the local orthorhombic strain distribution broadens strongly on
cooling. For a Gaussian distribution of local stresses that is weakly
temperature dependent, the nematic contribution to the relative
$d$-spacing width is
\begin{equation}
\begin{split}
\varepsilon_{{\rm nem},hkl}(T)
&=
\Gamma |g_{hkl}|
\frac{\Delta_{\sigma}}{C_{66}(T)}  \\
&=
B_{hkl}
+
\frac{A_{hkl}}{T-T_{s}^{\rm CW}},
\end{split}
\label{eq:fwhm_cw}
\end{equation}
where $\Gamma=2\sqrt{2\ln2}$ and
$g_{hkl}=\partial\ln d_{hkl}/\partial\varepsilon_6$ projects the
orthorhombic strain onto the selected Bragg reflection. For a pure
orthorhombic distortion in the axes used here, $g_{H00}=1$,
$g_{0K0}=-1$, and $g_{00L}=0$. Equation~(\ref{eq:fwhm_cw})
therefore explains why the in-plane reflections exhibit a strong
Curie--Weiss-like broadening while the $c$-axis width in
Fig.~\ref{fig3}(d) shows no comparable anomaly.

This expression also clarifies the meaning of the fitted temperature
scale $T^*$. In the ideal limit where the measured width is dominated
by the nematic-related contribution, the pole of the Curie--Weiss-like
fit corresponds to the coupled nematoelastic scale
$T_s^{\rm CW}$. In real samples, the measured width may also contain
instrumental resolution, compositional broadening, mosaicity, and
non-nematic microstrain. For statistically independent Gaussian
contributions, the variances add according to
\begin{equation}
\begin{split}
\varepsilon_{{\rm FWHM},hkl}^{2}(T)
&=
\varepsilon_{{\rm bg},hkl}^{2}(T)  \\
&+
\varepsilon_{{\rm nem},hkl}^{2}(T).
\end{split}
\label{eq:fwhm_background}
\end{equation}
Accordingly, the absolute value of $T^*$ can depend on the
background contribution and fitting range. However, the systematic
doping evolution of $T^*$, the strong contrast between in-plane and
$c$-axis responses, and the correspondence with elastoresistance in
{\BFCA} show that the temperature-dependent part of
$\varepsilon_{\rm FWHM}$ is governed by the nematoelastic lattice
response.

The {\BFCA} series provides the most direct test of this picture. With
increasing Co doping, both the magnitude of the in-plane
Curie--Weiss-like broadening and the fitted $T^*$ are systematically
suppressed. This mirrors the weakening of nematicity across the
electron-doped phase diagram. Elastoresistance and neutron Larmor
diffraction probe this instability in different ways. Elastoresistance
measures an electronic response to an imposed strain, whereas Larmor
diffraction measures the distribution of lattice spacings produced by
local stresses acting through the orthorhombic compliance. The close
parallel between the two doping trends therefore provides strong
evidence that the strain-free in-plane $d$-spacing distribution carries
quantitative information about the same underlying nematoelastic
response. The same framework also accounts for the Curie--Weiss-like broadening observed in {\BKFA} and {\FS}, showing that the nematoelastic lattice response is not restricted to electron-doped BaFe$_2$As$_2$ and does not require static long-range antiferromagnetic order.

The behavior of {\FT} highlights the role of the transition character.
For a strongly first-order magnetostructural transition, the gradual
growth of the nematoelastic response is cut off before the continuous
instability is reached. A minimal phenomenological description is
obtained by replacing the continuous nematic free energy with a
sixth-order expansion,
\begin{equation}
\begin{split}
f_{\rm FO}(\phi)
=&\,
\frac{R(T)}{2}\phi^{2}
-\frac{u_{\rm FO}}{4}\phi^{4}
+\frac{v}{6}\phi^{6}
-h\phi, \\
R(T)
=&\,
a(T-T_{s}^{\rm CW}) ,
\end{split}
\label{eq:first_order}
\end{equation}
where $u_{\rm FO},v>0$ and
$h=\lambda\overline{\sigma_6}/C_{66,0}$ is generated by the
spatially averaged uniaxial stress. At zero field, the transition
occurs before the Gaussian susceptibility can diverge. This provides a
natural explanation for the absence of an extended Curie--Weiss-like
regime in nominally strain-free {\FT}, as well as in
SrFe$_{1.97}$Ni$_{0.03}$As$_2$ \cite{lu2016impact}. Equation
(\ref{eq:first_order}) should be viewed as a phenomenological model
of the measured symmetry-breaking lattice channel, rather than a
microscopic description of the full bicollinear magnetic and
monoclinic structural transition in FeTe.

Under uniaxial pressure, the situation changes qualitatively. The
average stress acts as a conjugate field and biases the orthorhombic
response, while spatial variations of the applied load create a
distribution of local stresses. The measured width can then be written
phenomenologically as
\begin{equation}
\begin{split}
\varepsilon_{{\rm FWHM},hkl}^{2}(T,P)
&=
\varepsilon_{{\rm bg},hkl}^{2}  \\
&+
\left[
\Gamma |g_{hkl}|
S_{66}(T,h)\Delta_{\sigma}(P)
\right]^{2},
\end{split}
\label{eq:pressure_fwhm}
\end{equation}
where $S_{66}(T,h)$ is the differential orthorhombic compliance in
the presence of the conjugate field. This expression captures the main
features of the pressure-applied {\FT} data. The pressure makes the
underlying nematoelastic response visible by biasing the
symmetry-breaking channel, and the nonuniform component of the load is
amplified into a measurable $d$-spacing distribution. The larger width
along the pressure-applied direction follows naturally because
uniaxial stress contains both symmetry-preserving and orthorhombic
strain components. These contributions add along the compressed
direction and partially compensate along the transverse direction,
producing the observed anisotropic broadening. This behavior closely
parallels the pressure-induced broadening previously observed in
SrFe$_{1.97}$Ni$_{0.03}$As$_2$ \cite{lu2016impact}.

The resulting physical picture is therefore unified. In systems with
continuous or weakly first-order nematic transitions, weak local
symmetry-breaking fields are amplified by the soft orthorhombic
compliance, producing Curie--Weiss-like broadening of the in-plane
$d$-spacing distribution even when the average structure remains
tetragonal. In strongly first-order systems, this gradual enhancement
is suppressed in the absence of external stress, but can reappear when
uniaxial pressure acts as a conjugate field and introduces a
spatially distributed strain. Neutron Larmor diffraction thus probes
the full bulk distribution of nematic-related lattice distortions,
providing information complementary to macroscopic shear-modulus and
elastoresistance measurements, as well as to local and mesoscopic
structural probes such as pair-distribution-function analysis and
dark-field x-ray microscopy
\cite{frandsen2017local,frandsen2019quantitative,yay2026discovery}.

In summary, we have shown that the relative $d$-spacing width measured
by neutron Larmor diffraction provides a sensitive bulk probe of
nematic-related lattice broadening in FeSCs. In {\BFCA}, the
Curie--Weiss-like enhancement of the in-plane $\varepsilon_{\rm FWHM}$
and the systematic doping evolution of $T^*$ track the suppression of
nematicity with Co doping. The related behavior in {\BKFA} and {\FS}
supports the broader applicability of this approach, while the
contrasting response of {\FT} demonstrates how a strongly first-order
transition and externally applied pressure modify the measured
distribution. These results establish neutron Larmor diffraction as a
direct structural probe of nematoelastic lattice response and provide a
basis for connecting local lattice broadening, elastic softening, and
electronic nematicity in FeSCs.

\begin{acknowledgments}
The work at BNU is supported by the Scientific Research Innovation Capability Support Project for Young Faculty (ZYGXQNJSKYCXNLZCXM-M2) (X.L.). R.L. acknowledge financial support from the China Postdoctoral
Science Foundation (Grant No. 2025M773375) and Beijing Natural Science Foundation (Grant No. 1264069). Materials synthesis and neutron scattering research on iron-based superconductors at Rice are supported by the U.S. DOE, BES under Grant Nos. DE-SC0012311 and DE-SC0026179 (P.D.). Part of the materials characterization efforts at Rice is supported by the Robert A. Welch Foundation Grant No. C-1839 (P.D.). This research used resources at the High Flux Isotope Reactor, a DOE Office of Science User Facility operated by the Oak Ridge National Laboratory. The beam time was allocated to HB-1 on proposal number IPTS-32700.

\end{acknowledgments}
%-------------------------------------------------------------------------------------
%\begin{thebibliography}{}
%------------------------------------------------------------------------------------
%-------------------------------------------------------------------------------------
%\end{thebibliography}
%------------------------------------------------------------------------------------
% \bibliographystyle{apsrev4-2}
% \bibliography{FeSCs_revised}


\end{document}